\documentclass[slac_one]{revtex4}
\usepackage{graphicx}
\usepackage{fancyhdr}
\usepackage{atlasphysics}

\newcommand{\B}{\ensuremath{ B}\xspace}

\newcommand{\Kns}{\ensuremath{ K^{0*}}\xspace}

\newcommand{\Jpsipa}{\Jpsi}

\newcommand{\lumihigh}{\lowL\xspace }
\newcommand{\lumihihh}{\highL\xspace }

\newcommand{\Jpsitomm}{\ensuremath{ \Jpsi \ra \mumu }\xspace}

\newcommand{\Bst}{\ensuremath{ \Bs \ra \JpsiPhi }\xspace}

\newcommand{\JpsiPhi}{\ensuremath{ \Jpsi \phi }\xspace}

\newcommand{\Bsmumu}{\ensuremath{ \Bs \ra \mumu}\xspace}

\newcommand{\BmumuKns}{\ensuremath{ \Bd \ra \mumu \Kns }\xspace}
\newcommand{\Bplusmumuk}{\ensuremath{ B^{+} \ra \mumu \kplus }\xspace}
\newcommand{\Bplusmmks}{\ensuremath{ B^{+} \ra \mumu K^{+*} (\kshort \pi^{+} ) }\xspace}

\newcommand{\Bsmumuphi}{\ensuremath{ \Bs \ra \mumu \phi }\xspace}
\newcommand{\Lamzerobmumu}{\ensuremath{\Lamzerob \ra \mumu \Lamzero }\xspace}
\newcommand{\Afbe}{\ensuremath{A_{FB}}\xspace}
\newcommand{\bsdll}{\ensuremath{ b \ra s(d) l^+l^- \ }\xspace}

\newcommand{\BtoKns}{\ensuremath{ \Bd \ra \Jpsi \Kns }\xspace}
\newcommand{\BtoKnsn}{\ensuremath{ \Bd \ra \Jpsi \Kns}\xspace}
\newcommand{\BtoKpi}{\ensuremath{ \Bd \ra \Jpsi \kplus \piminus }\xspace}

\newcommand{\Lamzero}{\ensuremath{\Lambda^{0} }\xspace}
\newcommand{\Lamzerob}{\ensuremath{\Lambda^{0}_{b} }\xspace}

\begin{document}
\pagenumbering{arabic}
\title{Prospects for observing CP violation and rare decays at ATLAS and CMS} 

%

\author{Maria Smizanska}
\affiliation{Lancaster University, UK}
\author{\it For ATLAS and CMS collaborations}

\begin{abstract}
 Investigating B-hadron decays
represents an alternative approach to direct searches for physics 
beyond the Standard Model (BSM). 
ATLAS and CMS concentrate on B decays that can be registered
by a di-muon signature.  B-hadrons decaying to \Jpsitomm 
  will statistically dominate B-physics analyses 
  allowing high precision measurements, in particular  
   a test of BSM effects in the CP violation of  \Bst.  
In the so-called rare B-decay sector,  ATLAS and CMS will concentrate on a family of
semi-muonic exclusive channels, $b \ra s \mumu$, and on the purely muonic decay \Bsmumu.
After three years of LHC running at a luminosity of a few 
times \lumihigh (corresponding to  30~$fb^{-1}$), 
each of these two experiments can measure  the \Bsmumu signal with  
   3~$\sigma$ significance, assuming the Standard Model (SM) value for the decay probability.

\end{abstract}

\maketitle

\thispagestyle{fancy}

\section{INTRODUCTION}
 In the past 
  decade  BaBar, Belle, CDF and D0 made very precise measurements of 
   flavour and CP-violating phenomena.
   Whilst the analysis of the remaining data of these experiments may still push 
   the boundaries, no evidence of physics beyond the SM, 
   nor any evidence for CP violation other than that originating from the CKM mechanism,  
   has yet been found. At LHC, thanks to the large beauty 
   production cross-section and the 
   high luminosity of the machine, the sensitivity of B-decay measurements is 
   expected to substantially improve. Whilst direct detection of new particles 
   in ATLAS and CMS will be the main approach to discovering new physics,
   these two experiments will also perform B-Physics searches to 
   provide complementary information on BSM physics from  B-decays. 
 In the present document we concentrate 
    on three of the most prominent cases: the rare decay \Bsmumu, CP violation in \Bst and 
     the semi-muonic exclusive channels of type $b \ra s \mumu$.

\section{TRIGGERS FOR B-DECAYS WITH DI-MUON SIGNATURES} 
\begin{figure}[h]
\begin{center}
  \hfill
  \begin{minipage}[t]{0.59\textwidth}
       \includegraphics[width=8cm,height=4.5cm]{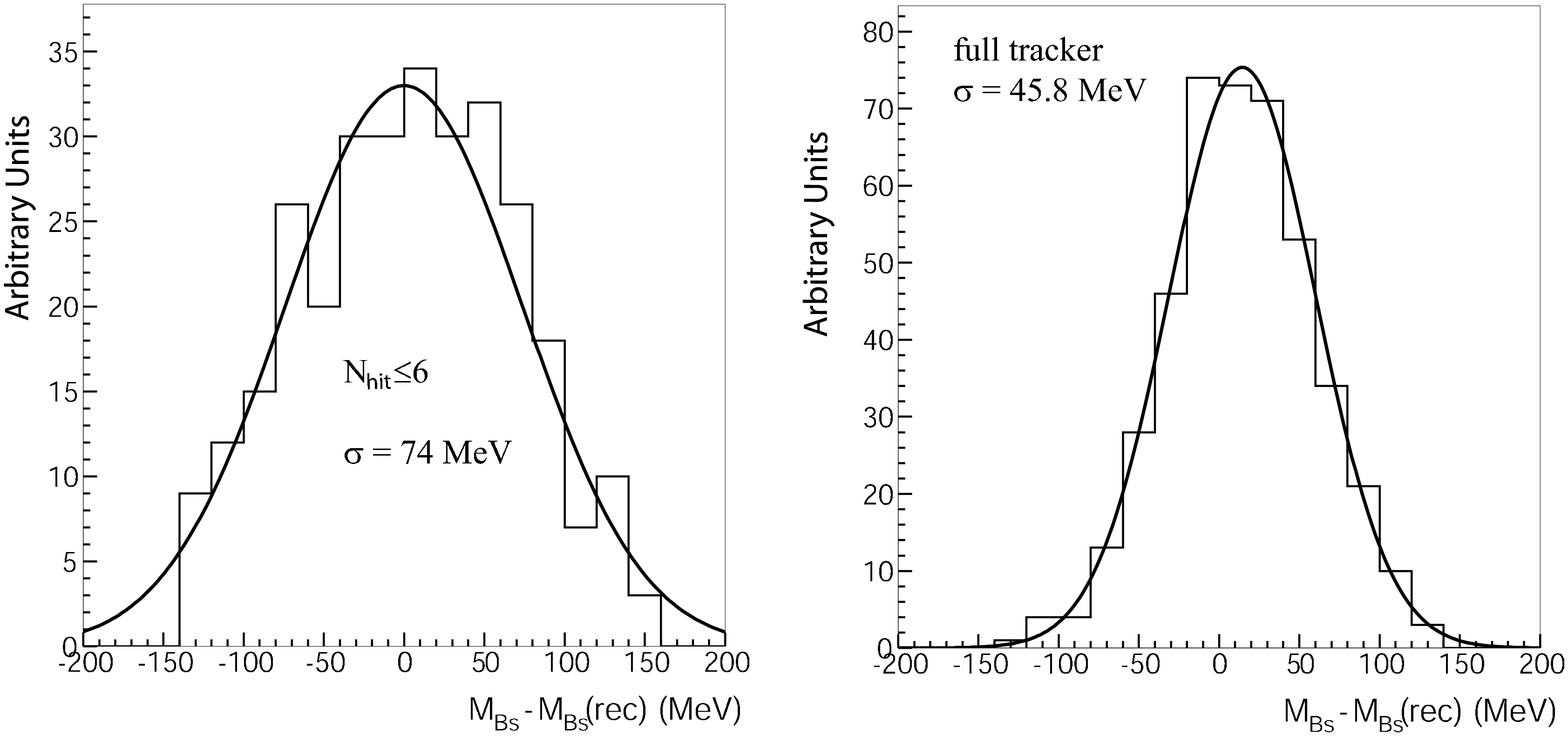}
\end{minipage}
  \hfill
  \begin{minipage}[t]{0.39\textwidth}
       \includegraphics[width=5cm,height=4.5cm]{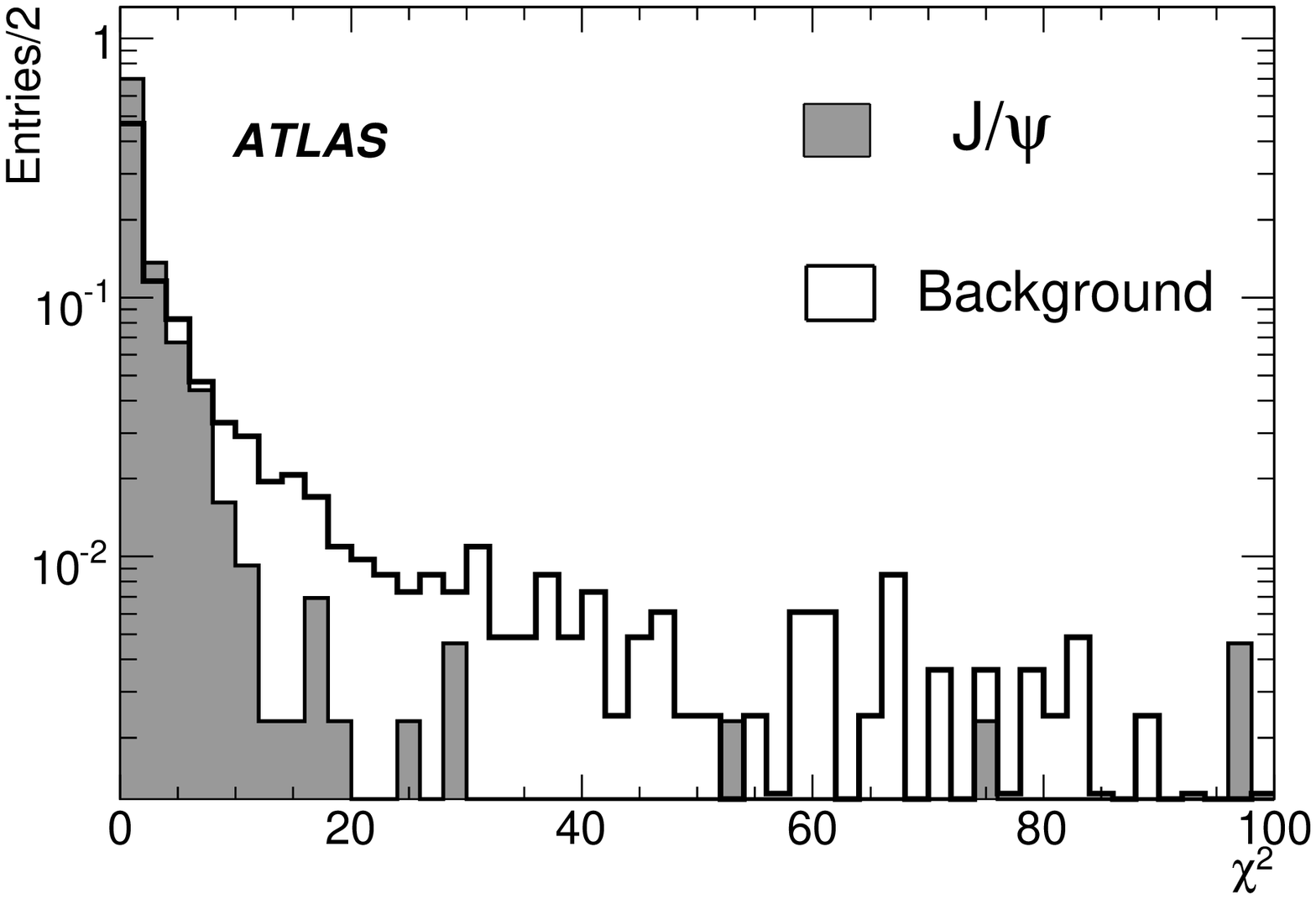}
  \end{minipage}
  \hfill
  \end{center}
  
\caption{Reconstruction of di-muon signals in the second-level trigger. 
 CMS   \Bsmumu mass reconstruction using fast L2 track reconstruction (left), 
 and the same for the offline reconstruction (middle).
 Right, L2 di-muon vertex fit ($\chi^2$) 
 for the \Jpsitomm signal and for di-muon background in ATLAS.}
 \label{fig:l2}
\end{figure}

 Both the ATLAS and CMS triggers comprise three levels and the selection of 
 B-decays with two muons in the final state is initiated by a 
 di-muon at the first trigger level (L1). 
At luminosity $10^{32}$~cm$^{-2}$~s$^{-1}$   the lowest possible threshold of about \pt $>$ 4~\GeV (3~\GeV) in ATLAS (CMS)
 will be used, rising to 6~-~8~\GeV at \lumihigh. 
 The muon is confirmed at the second trigger level (L2) combining 
 muon and inner detector (ID) information. In addition this trigger makes cuts on 
the invariant mass and certain secondary vertex 
qualities of the \B decay products to select specific physics processes. 
Figure \ref{fig:l2} illustrates the \Bsmumu mass resolution  
and the \Jpsitomm vertex reconstruction in the CMS and ATLAS L2 triggers respectively.
The channels  \Bst   and \Bsmumu are triggered by requiring di-muons fulfilling \Jpsipa or \Bs\xspace mass cuts. 
For semi-muonic rare channels, such as $ B_{d}{\rightarrow}K^{*\circ}\mu^+\mu^-$ 
or $ B_{s}{\rightarrow}\phi\mu^+\mu^-$, 
ID tracks are combined to first reconstruct the $K^{*\circ}$ or $\phi$; 
the muon tracks are then added to reconstruct the $\rm B$.
Selected  events  are further analysed at the third trigger level by algorithms 
using full event information. Approximately ten B-physics events 
will be recorded per second to permanent storage in each ATLAS and CMS.

\section{SENSITIVITY IN \Bsmumu} 
\begin{figure*}[t]
\centering
\includegraphics[width=50mm]{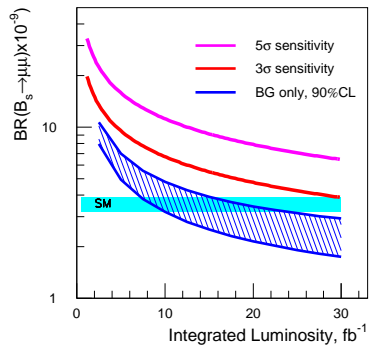}
\caption{ Upper  limits on the branching ratio 
of $B^{0}_{s}\to\mu^{+}\mu^{-}$ for each ATLAS and CMS.
  The results are given at confidence level 
of 90\% as a function of integrated luminosity.} 
\label{reach}
\end{figure*}

In the SM the di-muonic rare decay  \Bsmumu 
is mediated by flavour-changing neutral currents that are forbidden   
at tree level.  In addition this decay is helicity suppressed, resulting in an
expected branching ratio (BR) of $(3.42\pm0.52)\,\times\,10^{-9}$ ~\cite{Buch93Bur03}.
The motivation for precise measurements of rare B-decays is to test the SM 
 up to high perturbative orders, as well as to search 
for BSM effects that can increase the BR compared to SM predictions.
 After \Bsmumu events have passed the triggers the offline selection criteria  
are applied, see Table~I.
 With 30~$fb^{-1}$ (expected after three years of running at a 
 luminosity of a times of \lumihigh),  ATLAS and CMS will be able to select 
17.1~$\pm$~6  and  18.3~$\pm$~6.3 signal events respectively, assuming the SM value for the BR. 
Number of  $42^{+39}_{-30} $ (ATLAS) and   $42^{+66}_{-42}$ (CMS)
 background events $b\bar{b} \to \mu^+\mu^- X$ are expected. 
  The errors are dominated by limited statistics of simulated events. 
 Contributions from other backgrounds:  $B_s^0\to K^-\mu^+\nu$, $B_d^0\to K^-\pi^+$ and 
 $B_d^0\to \pi^-\pi^+$ are negligible, see \cite{CSC5} and \cite{CMSrare} for reference. 
These results were  used to estimate  upper 
limits on the branching ratio of $B^{0}_{s}\to\mu^{+}\mu^{-}$, 
shown in Figure~\ref{reach}.
Both ATLAS and CMS were using the algorithms of Ref.~\cite{Eidelman:2004wy}.
After three years at luminosity \lumihigh, both ATLAS and CMS can reconstruct the
\Bsmumu signal  with significance 3~$\sigma$.
  The strategy is to continue the di-muon channel programme up
to nominal LHC luminosity \lumihihh.

\begin{table}
\begin{tabular}{|l|c|c||l|c|c|}
\hline \hline

\multicolumn{3}{|c|}{  ATLAS } & \multicolumn{3}{||c|}{  CMS }  \\ \hline 
Selection cut     & $B_s^0\to\mu^+\mu^-$ & $b \bar{b} \to \mu^+\mu^- X$  &
Selection cut     & $B_s^0\to\mu^+\mu^-$  & $b \bar{b} \to \mu^+\mu^- X$ \\ \hline \hline

$I^{atl}_{\mu\mu} > 0.9$, $L_{xy} > 0.5$mm,       &     & 
&$I^{cms}_{\mu\mu} > 0.85$,$L_{xy} > 0.8$mm& &\\
$\alpha < 0.017$ rad,        &   $\epsilon = 0.04$  &  $\epsilon =0.24 \cdot 10^{-6}$ & 
$\alpha < 0.1$ rad & $\epsilon = 0.016 $ & $\epsilon =0.26 \cdot 10^{-6} $ \\ 

-70 MeV $< \Delta m < $140 MeV  & &   & $|\Delta m |< $ 100 MeV& & \\  

\hline \hline
Event yield  30~$fb^{-1}$ & 17.1 $\pm$ 6 &  $42^{+39}_{-30} $ &   & 18.3 $\pm$ 6.3 & $42^{+66}_{-42} $ \\ 
\hline \hline  
\end{tabular}
\caption{ Selection efficiencies, $\epsilon$,  and numbers of signal and background events for 
 30~fb$^{-1}$. Efficiencies were calculated after preselection criteria had been applied:
$ \rm  \pt^{\mu} > 6(4) GeV$ for the first and second muon (ATLAS), 
$ \rm  \pt^{\mu} > 3 GeV$ and $ \rm  \pt^{\mu\mu} > 5 GeV$ (CMS);  
 $4 \GeV < m(\mu\mu) < 7.3 \GeV$ (ATLAS) and
$5 \GeV < m(\mu\mu) < 6 \GeV$ (CMS). $\Delta m$ denotes a mass difference $m(\mu\mu)-M_{Bs}$.
Variables $I^{cms}_{\mu\mu}$, $I^{atl}_{\mu\mu}$ of di-muon isolation, as well as
 the angle $\alpha$ and the distance $L_{xy}$ have been  
 defined in \cite{CSC5} and \cite{CMSrare} for ATLAS and CMS respectively .}
\label{table:rare}
\end{table} 

\section{CP VIOLATION AND OTHER PARAMETERS OF \Bst decay}

New phenomena beyond the SM may alter CP violation in B-decays. A prominent 
example that has received experimental and theoretical attention is the decay 
 \Bst. Combined D0 and CDF results on a pair of parameters $-$ the
mixing phase $\phi_s$ and the decay width difference
  $ \Delta \Gamma_s$ $-$ yields a 2.2~$\sigma$ deviation from the SM 
 \cite{CDFD0}. 
The LHC is expected to produce the statistics necessary to unambiguously 
confirm or reject possible BSM contributions in this decay. 
With an integrated luminosity of 30~$fb^{-1}$ ATLAS and CMS will accumulate $2.4\cdot 10^{5}$ and $2.6\cdot 10^{5}$ \Bst events respectively. 
Simulation of the time dependent three-dimensional angular distribution 
of the final state particles proved that both experiments can perform  a  
model independent determination of $\phi_s$ and $ \Delta \Gamma_s$, along with  
 the average  width $\Gamma_s$ and some parameters of three  
 helicity amplitudes  of the decay. 
  The factors determining the measurement sensitivity $-$ the \Bs proper decay-time resolution, 
  flavour tag quality, signal statistics and background $-$ are 
 summarized in Table \ref{table:sumary}. 
 Under these assumptions the weak phase $\phi_s$ will be determined with a precision of 0.067 
 in both ATLAS and CMS. Whilst this precision is not enough to confirm the SM 
 estimate of $\phi_s$ (-0.0368~$\pm$~0.0018), it could verify a possible enhancement
  from BSM contributions as predicted  for instance in \cite{ball}. 
  The expected precisions of other parameters are given in Table \ref{table:sumary}.  
   The strong phases of the  helicity amplitudes have been found to be highly 
   correlated in the fits 
and it is thus supposed that these will be determined from the process \BtoKnsn. 
 
\begin{table} 
\begin{tabular}{|l|c|c||c|c|c|}
             \hline
                            &     ATLAS   &     CMS&  Statistical error  on parameter   & ATLAS & CMS \\   \hline\hline
                  Signal   statistics, 30~$fb^{-1}$    &  $ 2.4\cdot 10^{5}$           & $2.6\cdot 10^{5}$
 & $\delta \phi_{s}$              & $0.067 $  &  $0.067   $   \\
              Proper time resolution    &    $0.083 \mathrm ~ ps$           &   $0.077 \mathrm ~ps$ 
& $\delta \Delta \Gamma_{s}/{\Delta \Gamma_{s}}$    & $13\% $  &  $12\%$       \\
             \Bs mass resolution &  16.6 MeV    & 14 MeV  
& $\delta \Gamma_{s}/{ \Gamma_{s}}$    & $1\% $  &  $0.9\%$       \\  
              Background    &  $\sim 30\%$          & $\sim 33\%$      
& $ \delta A_{||}/A_{||}$ & $0.7\% $  &  $0.8\% $    \\
                       Tag quality using   $\mu$, e, jet-charge tags          &   $  3.9$          &  $  3.9$ 
&$\delta A_{\perp}/A_{\perp}$  & $3\% $  &  $2.7\% $      \\  \hline \hline

  \end{tabular}
           \caption{\Bst $-$ summary  of performance and expected statistical errors of  parameters  
          for an integrated luminosity of  30~$fb^{-1}$. Background is dominated by $\BtoKns, \BtoKpi$ decays. }        
           \label{table:sumary}   
 \end{table}

\section{PROGRAMME FOR RARE EXCLUSIVE SEMILEPTONIC  DECAYS}
\begin{figure}[h]
\begin{center}
  \hfill
  \begin{minipage}[t]{0.45\textwidth}
       \includegraphics[width=5cm,height=4.5cm]{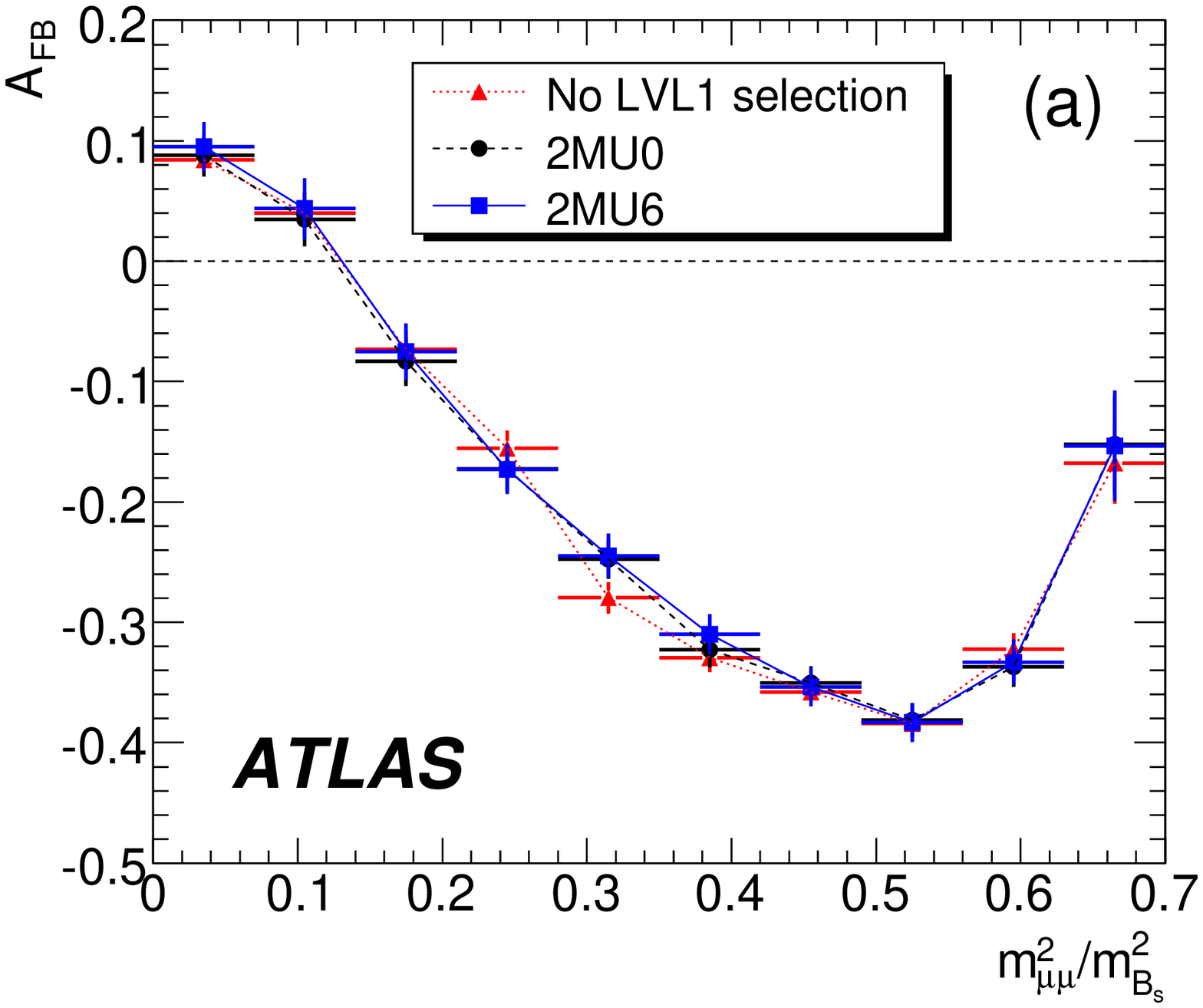}
\end{minipage}
  \hfill
  \begin{minipage}[t]{0.45\textwidth}
       \includegraphics[width=5cm,height=4.5cm]{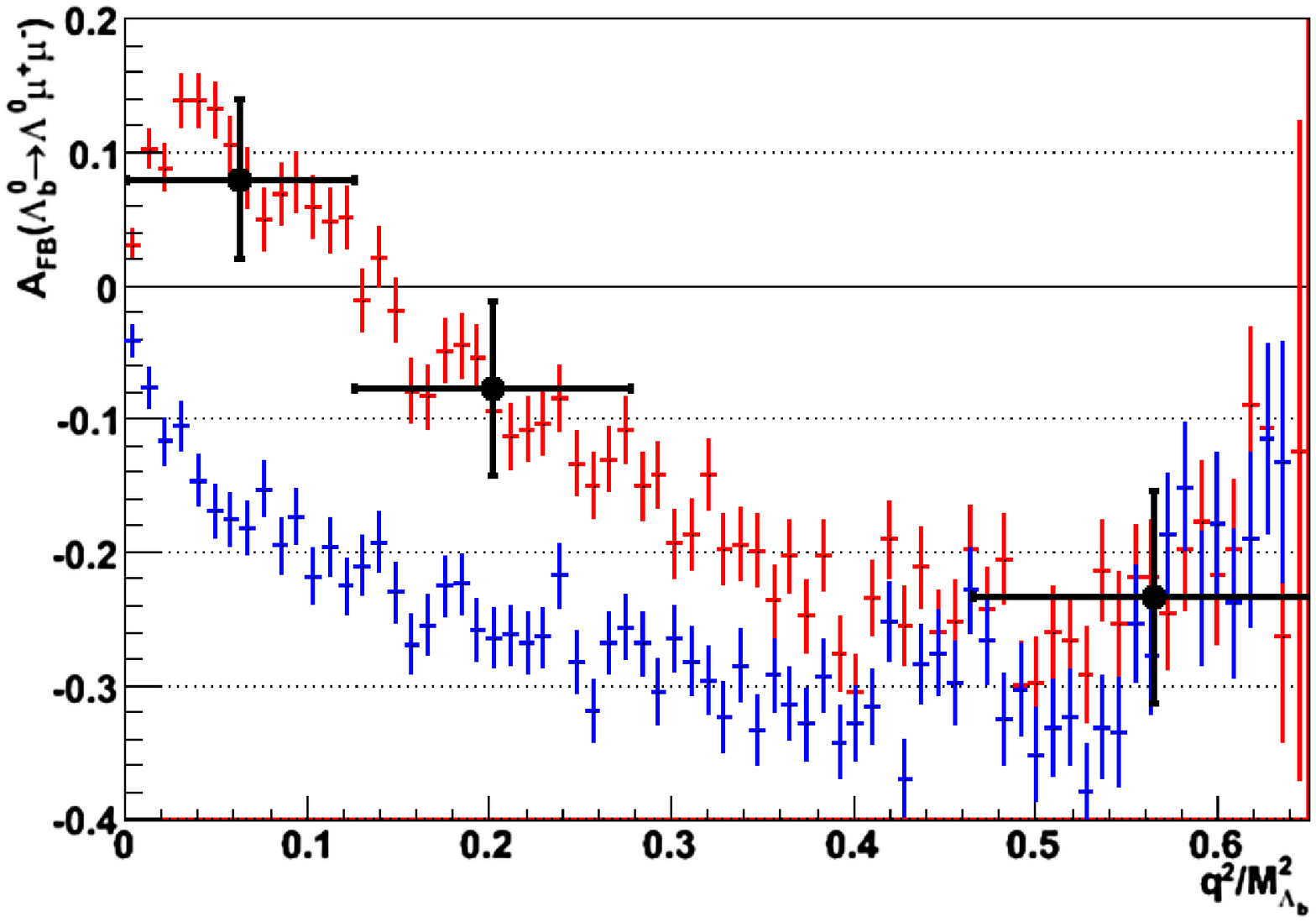}
  \end{minipage}
  \hfill
  \end{center}
  \caption{Left, \Afbe for \Bsmumuphi events in ATLAS: before L1 trigger (filled 
triangles), after di-muon L1 trigger with $\pt^{\mu} >$ 4~GeV (filled circles) 
and with $\pt^{\mu} >$ 6~GeV (open squares). Right, ATLAS sensitivity of \Afbe measurement
in \Lamzerobmumu, three large points with error bars show ATLAS sensitivity with 30 $fb^{-1}$, 
upper set of points generated by SM, lower set by MSSM model in Ref. \cite{xx} .}  
  \label{fig:semi}
\end{figure}

 In the SM, the electroweak penguin decays \bsdll  are 
 only induced at the one-loop level, leading to small branching 
fractions and thus a sensitivity to contributions from BSM. 
 Problematic hadronic form-factor uncertainties entering 
 the calculations of branching ratios lead physicists to observables 
 that are less unaffected by the uncertainties and are therefore a better basis 
 for BSM searches.  The most frequently used  
 forward~$-$~backward asymmetry \Afbe is defined through the angle  
 between the $\mu^+$ and the B-hadron momenta in di-muon rest frame \cite{refAFB}.
In ATLAS the shape of the \Afbe distribution as a function of the di-muon mass is negligibly affected by
applying di-muon triggers (Fig. \ref{fig:semi}). 
This feature will make the results only weakly  
dependent on trigger acceptance calibrations \cite{CSC5}. 
A list of decays supported by ATLAS is given in Table \ref{tabll}  
together with the corresponding statistics after offline selections have  
reduced the background. The background determination  
was limited by MC statistics and therefore only an upper limit is given. 
Sensitivity studies were performed for each channel and in particular 
for \Lamzerob decay a precision of 6\% on the \Afbe  can be 
achieved after 30 $fb^{-1}$, see Fig. \ref{fig:semi}. 
This will allow to distinguish the SM from one example of MSSM, calculated for \Lamzerobmumu in Ref. \cite{xx}

\begin{table} 
\begin{tabular}{|l|c|c|l|c|c|}

\hline
Process & Signal & $b \bar{b} \to \mu^+\mu^- X$  &Process & Signal & $b \bar{b} \to \mu^+\mu^- X$  \\
        \hline\hline           
\BmumuKns             &  2500      &  $<$ 10000 &\Bsmumuphi & 900  & $<$ 10000   \\  
\Bplusmumuk & 2300 & $<$ 10000 &\Lamzerobmumu & 800& $<$ 4000 \\ 
\Bplusmmks & 4000 & $<$ 10000 &&& \\ 
\hline \hline
 \end{tabular}  
\caption{Signal and background statistics in semileptonic rare decays in ATLAS.} 

\label{tabll}
\end{table} 

\section{CONCLUSIONS}
The ATLAS and CMS trigger and offline tools for di-muon B-physics channels 
are well prepared for the luminosities up to few times of \lumihigh. 
Both experiments are ready to perform measurements which will be 
sufficiently precise to allow confirmation of some possible BSM physics contributions.


\begin{thebibliography}{9}   

\bibitem{Buch93Bur03} G. Buchalla and A. J. Buras, Nucl. Phys. 
B400, 225 (1993); A.J. Buras, Phys. Lett. B566, 115 (2003).
\bibitem{CSC5} ATLAS Collaboration, 
  ``Expected Performance of the ATLAS Experiment,
  Detector, Trigger and Physics'',
  CERN-OPEN-2008-020, Geneva, 2008, to appear.
\bibitem{CMSrare} C. Eggel, U. Langenegger and A. Starodumov, CMS-CR-2006-071.
\bibitem{Eidelman:2004wy}
  S.~Eidelman {\it et al.}  [Particle Data Group],
  Phys.\ Lett.\  B {\bf 592} (2004) 1.
 \bibitem{CDFD0} D. Strom for CDF and D0 collaborations,
  "Bs parameters in \Bst", current proceedings.
\bibitem{ball} P. Ball "Probing new physics through B(s) mixing", 
 42nd Rencontres de Moriond on Electroweak Interactions and Unified Theories,
  La Thuile, 10-17 Mar 2007, e-Print: hep-ph/0703214.  
 \bibitem{refAFB} 
C.~Bobeth, G.~Hiller, G.~Piranishvili, JHEP 0712:040 (2007). 
  \bibitem{xx} T.~M.~Aliev, A.~Ozpineci and M.~Savci, Nucl.Phys.B 649 (2003) 168.  


\end{thebibliography}
\end{document}